# Arctic Oscillation: possible trigger of COVID-19 outbreak

Arturo Sanchez-Lorenzo[a*], Javier Vaquero-Martínez[a], Joan-A. Lopez-Bustins[b;] Josep Calbó[c],

Martin Wild[d], Ana Santurtún[e], Doris Folini[d], José-M. Vaquero[a], Manuel Antón[a]

[a] Department of Physics, University of Extremadura, Badajoz, Spain

[b] Climatology Group, Department of Geography, University of Barcelona, Barcelona, Spain

[c] Department of Physics, University of Girona, Girona, Spain

[d] Institute for Atmosphere and Climate (IAC), ETH Zurich, Zurich, Switzerland

[e] Unit of Legal Medicine, Department of Physiology and Pharmacology, University of Cantabria, Santander, Spain

**Abstract**: The current COVID-19 pandemic is having detrimental consequences worldwide. The pandemic started to develop strongly by the end of January and beginning of February 2020, first in China with subsequent rapid spread to other countries with new epicenters of the outbreaks concentrated mainly within the 30-50ºN latitudinal band (e.g., South Korea, Japan, Iran, Italy, Spain). Simultaneously, an unusual persistent anticyclonic situation prevailing at latitudes around 40ºN was observed on global scale, in line with an anomalously strong positive phase of the Arctic Oscillation. This atypical situation could have resulted in favorable meteorological conditions for a quicker spread of the virus over the latitude band detailed above. This possible connection needs further attention in order to understand the meteorological and climatological factors related to the COVID-19 outbreak, and for anticipating the spatio-temporal distribution of possible future pandemics.

The world is currently undergoing the COVID-19 pandemic associated with the SARS-CoV-2 coronavirus (Zhou et al., 2020). Several studies have shown that some meteorological variables, especially air temperature and humidity, can play a role in the spread of COVID-19 (Araujo and Naimi, 2020; Bu et al., 2020; Chen et al., 2020; Luo et al., 2020; Notari, 2020; Sajadi et al., 2020; Wang et al., 2020), even though no evidence of a major role of air temperatures has also been reported (Jamil et al., 2020; Martinez-Alvarez et al., 2020). Overall, dry conditions with mild temperatures seem to increase the transmission of the disease, in line with other respiratory viruses like influenza (Kudo et al., 2019).

Interestingly, it has been pointed out recently that the first main COVID-19 outbreaks occurred in regions and countries along an east-west longitudinal direction within a latitude band between 30-50ºN. On average, it seems that these regions had in common a mean air temperature of around 5-11ºC and low specific humidity levels of around 3-6 g/kg (Sajadi et al., 2020). Thus, after the outbreak in China in early January 2020, the disease mainly moved in February and March to countries such as Japan, South Korea, Iran, Italy and Spain, which are all located within the above mentioned latitudinal band, while not affecting significantly other regions immediately north and south of China. This spatial spread of the main outbreaks between January and early March 2020 does not seem to follow either population proximity connections or volume of international exchanges (Sajadi et al., 2020). Therefore, it is plausible that the anomalous atmospheric circulation pattern that occurred between January and March 2020, characterized by strong high-pressure anomalies centered in the 40ºN belt, helped to modulate the spatial pattern of the first outbreaks of the COVID-19.

Figure 1 (left) shows the global distribution of the geopotential height anomalies from January to March (JFM) 2020 for the 300 hPa (top) and 500 hPa (bottom) levels as derived from the NCEP/NCAR DOE Reanalysis product (Kanamitsu et al., 2002). A clear evidence of a stable atmospheric circulation can be seen (i.e., positive anomalies) along the 30-50ºN belt of the Northern Hemisphere, and the opposite (i.e., negative anomalies) north of 60ºN. The latitudinal averages of the 300 hPa and 500 hPa (Figure 1, right) highlight a maximum of the anomalies centered just at 40ºN, whereas very negative anomalies are reached above 70ºN. This spatial pattern fits with the known features related to the positive phase of the Arctic Oscillation (AO).

The AO is a teleconnection pattern that summarizes the northern polar vortex variability at surface level (Baldwin et al., 2003), and is characterized by pressure anomalies of one sign over the Arctic and the Norwegian sea, balanced by anomalies of opposite sign centred around latitudes between 35 and 50° of the Northern Hemisphere. The positive phase pattern of the AO shows lower-than-average air pressure over the Arctic region and higher-than-average pressure at mid-latitudes. The AO index for JFM 2020 reached exceptionally high values during these three months; in fact, they are the maximum ever recorded for the available 1950-2020 period (Figure 2). These anomalous positive values of the AO resulted in a reinforced polar vortex during JFM 2020, in line with the maps shown in Figure 1. A strong polar vortex in February and March delays the spring arrival in the polar stratosphere, keeping very low temperatures in its core and producing stratospheric ozone depletion (Rex et al., 2004). Indeed, a remarkable ozone loss was registered over the Arctic region during last March 2020 (Witze, 2020). Moreover, it is interesting to note that the dynamical forcing by the greenhouse gases related to the anthropogenic climate change could lead to an enhancement of the AO (Shindell et al., 2001).

It seems reasonable that the anticyclonic belt conditions favoured, to some extent, the spread of the virus. In fact, for example, the rate of deaths per population by the end of March 2020 averaged over latitude bands shows a clear maximum centred around 40ºN, which matches the peak in positive air pressure anomalies (Figure 1, right). It can be argued that along this band corridor around 35-50ºN, the positive pressure anomalies caused a tendency towards dry and calm conditions, as well as temperate conditions, that fit with the optimal meteorological conditions for the initial spread of the COVID-19 as has been reported by Sajadi et al. (2020). For instance, Sanchez-Lorenzo et al. (2020) have shown that the unusual persistent anticyclonic situation prevailing in southwestern Europe during February 2020 could have resulted in ideal atmospheric conditions for a quicker spread of the virus compared with the rest of the European countries. Therefore, in a year without this anomalous positive AO index, the same mild temperatures and low specific humidity levels recorded this JFM 2020 might have not been observed, and consequently a different spatio-temporal spread of the COVID-19 disease would have been expected.

To summarize, we highlight that an anomalous atmospheric circulation in early 2020, characterized by a half-century record AO's positive phase, may partially explain the first steps in the spread of the COVID-19, which tended to concentrate along a latitudinal band centered at 40ºN. This anomalous atmospheric circulation may have provided optimal meteorological conditions for the virus propagation by decreasing the body's defense mechanisms, favoring the social contact patterns and/or enhancing the transmission of the virus by means of the droplets and aerosols emitted after a cough. In fact, the strong stability associated with the anticyclonic conditions and subsidence along the 40ºN band may have promoted the airborne transmission, which has been recently suggested as a crucial way to infect people with the SARS-CoV-2 coronavirus (Morawska and Cao, 2020). Equally, the tendency towards atmospheric stability can limit the dispersion of the natural and anthropogenic pollutants, and high pollution levels have also been related to an increase in the incidence of COVID-19 fatalities (Ogen, 2020; Wu et al., 2020). Overall, the possible connection of the large-scale circulation on global scale and the COVID-19 outbreaks needs further research to better understand the physical and biological mechanisms, as well as the potential spread of other respiratory-origin pandemics in the past and future.

**Acknowledgments**

A. Sanchez-Lorenzo was supported by a fellowship RYC-2016–20784 funded by the Ministry of Science and Innovation. Javier Vaquero-Martinez was supported by a predoctoral fellowship (PD18029) from Junta de Extremadura and European Social Fund. J.A. Lopez-Bustins was supported by Climatology Group of the University of Barcelona (2017 SGR 1362, Catalan Government) and the CLICES project (CGL2017-83866-C3-2-R, AEI/FEDER, UE). This research was supported by the Economy and Infrastructure

Counselling of the Junta of Extremadura through grant GR18097 (co-financed by the European Regional Development Fund). Juan V., Xavi B. and Raúl J.I. (UPV/EHU) kindly helped us in discussing the results. NCEP Reanalysis data provided by the NOAA/OAR/ESRL PSL, Boulder, Colorado, USA, from their Web site at https://psl.noaa.gov/

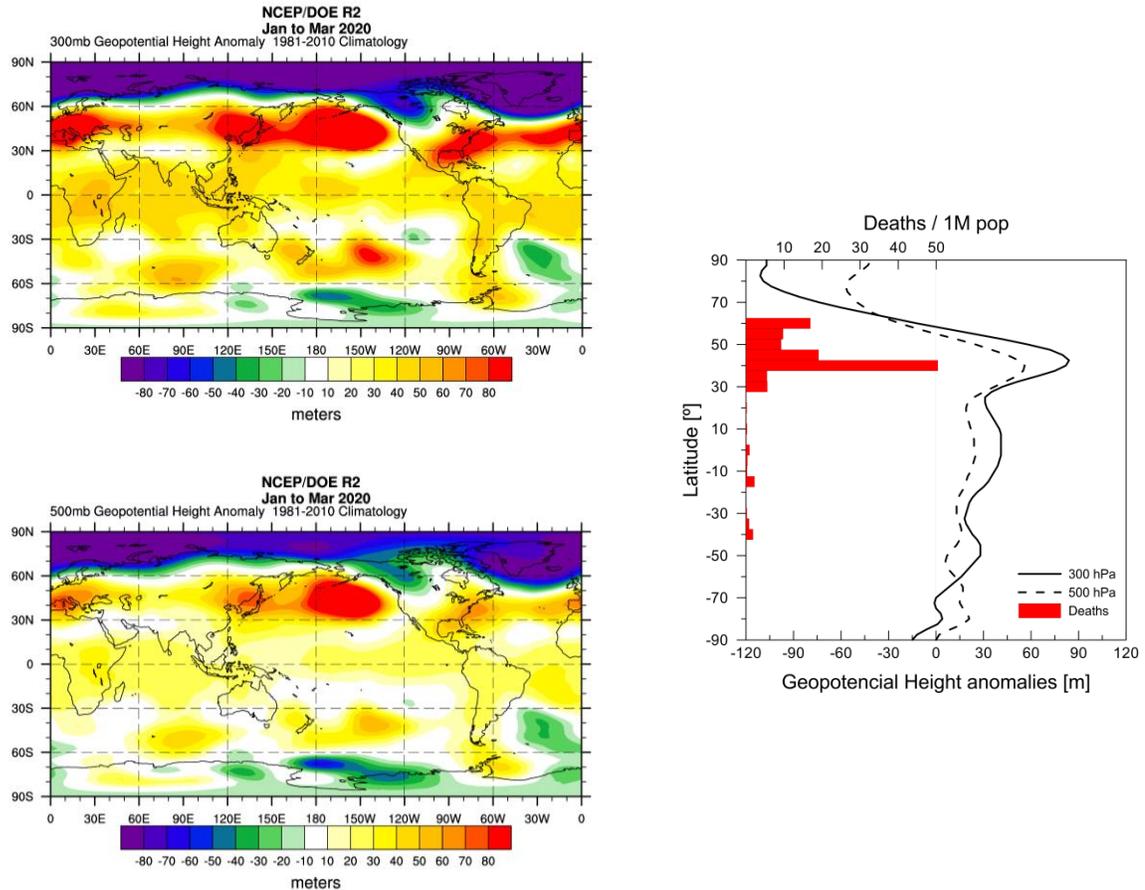

Figure 1. (left) Anomaly maps of (top) 300 hPa and (bottom) 500 hPa geopotential heights (m) for JFM 2020 over the World. (right) Latitudinal distribution of the 300 hPa (dashed line) and 500 hPa (solid line) geopotential height (m) anomalies, together with the number of officially reported COVID-19 deaths per million of inhabitants in latitudinal bands of 5°. Anomalies are calculated with respect to the climatological mean (1981-2010 period). NCEP/DOE R2 reanalysis data is used here, with a spatial resolution of 2.5° per latitude and longitude. COVID-19 data on country basis were obtained on March 31[th], 2020 from the website https://github.com/CSSEGISandData, which it is provided by the Center for Systems Science and Engineering (CSSE) at the Johns Hopkins University. Population data have been obtained from the Center for International Earth Science Information Network - CIESIN - Columbia University (2018), Version 4 (GPWv4).

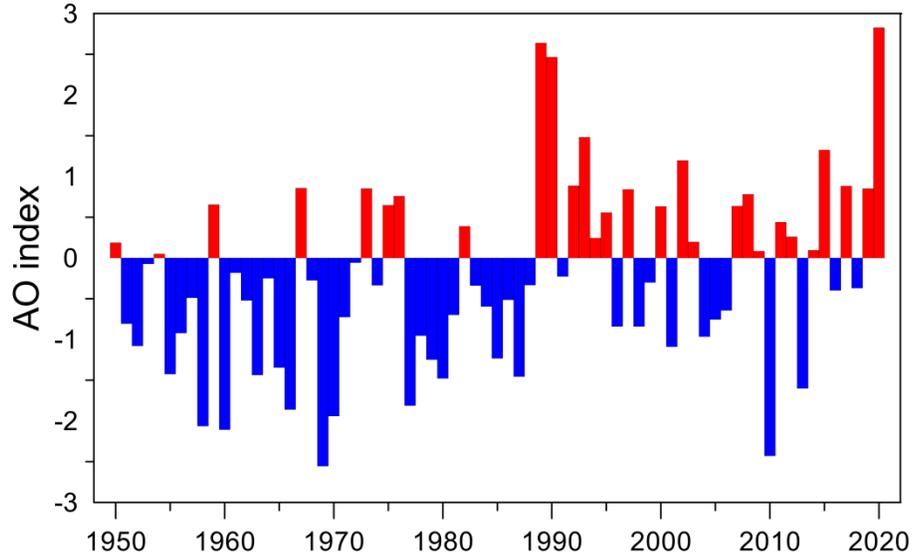

Figure 2. Average time series of the Artic Oscillation (AO) index from January to March (1951-2020 period). The JFM 2020 is the highest value (2.826), followed by 1989 (2.638) and 1990 (2.464) and far away from the 4th largest value in 1993 (1.481). The AO index has been extracted from the Climate Prediction Center of the National Oceanic and Atmospheric Administration (NOAA).